\documentclass[useAMS,usenatbib]{emulateapj}

\usepackage{epsfig}
\usepackage{amsmath}
\usepackage{natbib}
\usepackage{epstopdf}

\begin{document}

\title{
Local Circumnuclear Magnetar Solution to Extragalactic Fast Radio Bursts
}
\author{
Ue-Li  Pen$^{1,2}$
Liam Connor$^{1, 3, 4}$
}
\date{\today}

%
%

\begin{abstract}
We synthesize the known information about Fast Radio Bursts and radio magnetars, and
describe an allowed origin near nuclei of external, but non-cosmological, galaxies. 
This places them at $z\ll1$, within a few hundred megaparsecs.
In this scenario, the high DM is dominated by the environment of the FRB,
modelled on the known properties of the Milky Way Center, whose innermost 
100pc provides 1000 pc/cm$^3$. 
A radio loud magnetar is known to exist in our galactic centre, within $\sim$2 arc seconds
of Sgr A*.  
Based on the polarization, DM, and scattering properties of this 
known magnetar, we extrapolate its properties to those of Crab-like giant pulses and
SGR flares and point out their consistency with observed Fast Radio Bursts.
We conclude galactic center magnetars could be the source of FRB's.
This scenario is readily testable with VLBI measurements as well 
as with flux count statistics from large surveys such as CHIME or
UTMOST.
\end{abstract}

\footnote{$^1$ Canadian Institute for Theoretical Astrophysics, University of Toronto, M5S 3H8 Ontario, Canada} \\
\footnote{$^2$ Canadian Institute for Advanced Research, Program in Cosmology
and Gravitation}\\
\footnote{$^3$ Department of Astronomy and Astrophysics, University of Toronto,
M5S 3H8 Ontario, Canada}\\
\footnote{$^3$ Dunlap Institute for Astronomy and Astrophysics, University of Toronto, M5S 3H8 Ontario, Canada}

\newcommand{\be}{\begin{eqnarray}}
\newcommand{\ee}{\end{eqnarray}}
\newcommand{\beq}{\begin{equation}}
\newcommand{\eeq}{\end{equation}}

\section{Introduction}

The phenomenon of Fast Radio
Bursts (FRB's) has generated excitement in the astronomical community 
as well as speculation
regarding the origin of the events  
\citep{2007Sci...318..777L, 2013Sci...341...53T}. FRB's are millisecond radio transients 
with flux densities between 0.2$-$1.5 Jy. They are also highly dispersed, 
with DM's far exceeding the expected contribution from our own galaxy
in their direction
(DM$\sim$500$-$1200 pc/cm$^3$).
Theories of their distance vary from atmospheric to solar system, 
galactic and cosmological, however the
high DM's have lead a number of people to believe they are extragalactic. 
This is also partly due to their location on the sky, since the high 
galactic latitudes cast doubt on a galactic or solar
system origin
\citep{2013Sci...341...53T}.  
If the extragalactic dispersion is caused by the 
intergalactic medium (IGM) then the sources would be cosmological, found 
between redshifts 0.45$-$1. However it is possible that the dispersion 
could be due to dense regions in more nearby galaxies, as noted by 
\cite{2013Sci...341...53T} and \cite{2014ApJ...785L..26L}. These galaxies
would be within a few hundred megaparsecs, which we will consider 
non-cosmological.

Perhaps more mysterious than their location are their progenitors and 
emission mechanism. A wide range of ideas have been proposed, from 
Blitzars \citep{2014A&A...562A.137F}
to superconducting cosmic strings \citep{2014JCAP...11..040Y},
compact object mergers \citep{2013ApJ...776L..39K}) to nearby flaring main
sequence stars \citep{2014MNRAS.439L..46L}.

In this letter we provide yet one more allowed interpretation consistent with
current data: giant pulses or outbursts from magnetars 
in the nuclear regions of external galaxies.  The idea that
FRB's could be radio-emitting magnetars was explored by \cite{2014MNRAS.442L...9L}
and \cite{2014ApJ...797...70K}, where the high brightness 
temperatures are explained by shock-induced maser emission.
In our letter we do not focus on the emission mechanism nor do we favour 
pulsar-like emission (giant pulses) vs. SGR flares; we are simply putting forth 
an explanation for FRB's based
on magnetars near the centers of external galaxies that is consistent with the
 existing data, which makes falsifiable predictions. 
The non-cosmological (i.e. local) 
extragalactic nuclear FRB can naturally explain 
the observed large dispersion measures (DM) and scattering 
(SM) and could help explain their polarization properties.

\section{Galactic Center Pulsars}

\subsection{Nuclear Properties}

Our own galactic center region has a high measured electron density,
and the recently discovered pulsar and magnetar SGR J1745-2900 has a
measured DM=1778 \citep{2013Natur.501..391E}, most of which is thought
to originate from the inner few parsecs of the galaxy.  Seen from a typical
extragalactic viewing angle, this magnetar would have a DM $\sim$1000.
It is scattered by a few seconds at $\sim$ GHz, which is a thousand
times longer than the observed scattering time scales of FRB's. However VLBI
measurements indicate that this scattering is dominated by a screen
closer to our sun than the galactic center \citep{2014ApJ...780L...2B},
in which case a typical extragalactic line of sight would see a much
smaller scattering time, perhaps a few ms.

It had been thought that the GC harboured a large number of pulsars 
but that they were difficult to observe at low frequencies due to a scattering
screen within $\sim$200 pc of Sgr A* \citep{2012ApJ...753..108W, 1998ApJ...505..715L}.  
However after the discovery of the radio-loud magnetar J1745-2900 just 
$\sim$2 arcseconds from Sgr A* it now seems there really is a dearth 
of regular pulsars and an over-representation of magnetars. Though hundreds 
to thousands of ordinary pulsars were predicted to exists within $\sim.02$ pc of the 
galactic centre, none has yet to to be found \citep{2004ApJ...615..253P}.
\cite{2014ApJ...783L...7D} show that this implies the region is an effective
environment for magnetar formation, whose short lives could explain the lack
of such radio-loud objects in the central parsec. The GC would then be a
graveyard of highly-magnetized massive stars, some of which became 
magnetars and emitted in the radio for $\sim10^4$ years before 
spinning down sufficiently to cross the death line. 


It is worth pointing 
out that J1745-2900 is one of just four known radio-loud magnetars.
Within 2.2 arcseconds of Sgr A* it occupies volume that is
$\sim10^{-9}$ of our galaxy's volume and $\sim$5$\times 10^{-5}$ of 
its mass, and where there is an anomalous absence of ordinary pulsars. This
suggests to us that radio-loud magnetars not only $can$ form in such
environments, but preferentially do. External nuclear regions could therefore 
also harbour magnetars and could provide both the dispersion and the 
scattering observed in FRB's.

In the cosmological picture it is difficult to explain the scattering 
tails seen in several bursts from the IGM. \cite{2014ApJ...785L..26L} point out that 
if it is due to turbulence then the length scale 
of plasma scattering in the IGM at a distance of 1 Gpc for $\sim$ms tails
is impossibly small. In other models the IGM is an equally unlikely place 
for the scattering to occur \citep{2014MNRAS.442.3338P}. However we
do point out that \cite{2014ApJ...780L..33M} have shown cosmological FRB's
could be scattered at $\sim$ms by intervening galactic disks if 
their electron distribution were more extended than is currently believed.

\subsection{Possible Sources}

A dozen or so pulsars are known to exhibit giant pulses (GP's), which are 
of very short duration and can be many orders of magnitude brighter than their
average pulse flux \citep{2012ApJ...760...64M}.  A rare tail of {\it
  supergiant} pulses \citep{2004ApJ...612..375C} has also been
identified, with brightness 
temperatures reaching up to $10^{32-37}$K \citep{2003Natur.422..141H}.
They tend to 
be short enough ($<$16 ns) that their pulse is consistent with a pure
scattering profile, which is also the case for the observed FRB's.
It is worth noting
that the only FRB for which there is polarization information is 
FRB 140514, which was found to have $\sim$20\% circular polarization 
and very little linear polarization $<10\%$ \citep{2014arXiv1412.0342P}.
Considering the rotation
measure of the GC magnetar, J1745-2900, is RM=-6.7$\times$10$^4$ rad m$^{-2}$, 
if other galactic centers were like our own then nuclear pulsars and 
magnetars could become linearly depolarized due to multi-path 
Faraday rotation from a scattering screen \citep{2014arXiv1412.0342P}.
It is also possible that the sources themselves 
are circularly polarized.  At 2 GHz, J1745-2900 is also observed to be
$\sim$20\% circularly polarized, with no detected linear
polarization.  At higher frequencies, this magnetar is strongly
linearly polarized. Giant pulses are known to often be highly 
circularly polarized, for example over half of the peaks from B1937+21 
are in a pure Stokes V state. 

Though none of the known pulsars that exhibits GP's is a radio loud 
magnetar, the energetics of FRB's are not difficult to accommodate and 
one could
imagine high luminosity radio outbursts from such objects; some magnetars are 
soft gamma ray repeaters (SGR), which have episodic  outbursts emitting
$10^{46}$ erg 
in a fraction of a second.  At distances of $\sim$100 Mpc, the inferred
energy of an FRB is $\sim10^{36}$ erg, a tiny fraction of known SGR burst
energies. Magnetars that emit in the radio can also have non-negligible 
circular polarization and the GC object J1745-2900 seems to have a typical 
circular fraction of $20\%$, though this increases when the pulsar flares up
\citep{2014arXiv1412.0610L}. 

Only a tiny fraction of the burst energy needs to come out
to power a fast radio burst.  In order to explain the common large DM
of FRB's, these GP's would have to be preferential properties of
circumnuclear magnetars.  Events would be expected to repeat after
years, making a direct search challenging.  An all sky search with a
telescope such as CHIME \citep{2014SPIE.9145E..22B} 
over a year could discover $\sim10^5$ events,
of which $\sim10^4$ would repeat in a year 
and a few would be lucky enough to be caught in the
same CHIME beam a second time.  The long integrations at known FRB
locations have not resulted in repeat events, which is consistent with
this picture.

Given the small number of radio loud magnetars in the Milky Way, one 
cannot comment on their distribution in other galaxies. However in this 
picture there could be a sizeable fraction of sources that exist outside of 
their galaxy's nuclear regions, in which case there should be a 
commensurate number of FRB's with modest dispersion measures, 
perhaps 70$-$100 pc cm$^{-3}$ for an object at 100 Mpc. The apparent 
lack of sources with such DMs could be explained by a selection effect: 
radio bursts whose 
dispersion measures are not extraordinary may simply not get identified 
as FRB's. These may be missed or ignored given the large ensemble of 
radio transients with an apparent $\nu^{-2}$ sweep, including RRAT's
and perytons \citep{2012MNRAS.425.2501B}.

\section{Predictions}

This scenario is readily testable: 
At redshifts less than unity $z\ll 1$, the flux distribution is
given by a Euclidean universe, with $N(>S) \propto S^{-3/2}$, only
weakly dependent on DM, assuming the bursts are standard candle-like. 
This is not necessarily expected for high redshift
objects, where cosmological expansion and source population evolution
are expected to change. 

A VLBI detection would find a spatial coincidence to within a few
parsecs of a galactic nucleus, which is $\sim$ milli arcseconds at
distances of $\sim$100 Mpc.  The current non-coincidence with nearyby
galaxies constrains the typical distance to be larger than $\sim$100
Mpc.  This is still an order of magnitude closer than if the DM is
primarily accounted for by the intergalactic medium.

The galactic center magnetar is linearly depolarized at frequencies
below $\sim$4 GHz, consistent with multi-path Faraday depolarization
from the scattering screen \citep{2014arXiv1412.0342P}.  
Circular polarization is not affected, and
has indeed been observed in FRB's.

\section{Applications}

Substantial interest has developed for cosmology, should FRB be at
cosmological distances.  These are summarized in
\citep{2014ApJ...780L..33M}.  Should the DM be dominated by the host
galaxy, these applications would be difficult to materialize. The
expected scattering size of such events would be micro arcseconds,
which could be detectable with galactic
scintillation\citep{2014MNRAS.440L..36P}.

In a large survey, such as CHIME, the closest event could be at $\sim$
Mpcs distances.  Continuous
monitoring of neighboring galactic centers, e.g. M31, for years, could
detect pulses many kJy bright, requiring only a small receiver to
monitor.  Similarly, long term continuous monitoring of the GC
magnetar may uncover rare super-giant pulses.  All-sky telescopes,
such as the FFTT \citep{2009PhRvD..79h3530T}, may be well suited for
finding close, bright, sources.

Extrapolating from the one known nuclear magnetar, a sample of $10^4$
as might be found by CHIME, could result in the closest projected impact angle
of $\sim$7 mas.  This would place it near the Einstein Ring radius,
within $\sim 1000$ Schwarzschild radii, 
such that it could be gravitationally lensed by the central black
hole.  Assuming its projected proximity to the black hole does not increase 
the FRB's DM or SM too significantly,
this would be seen as an echo separated by the black hole
Schwarzschild time, $\sim$ seconds.  The echo would be fainter, and
the combination of delay and flux constrains the central black hole
mass.

\section{Conclusions}

We have described a FRB scenario based on circumnuclear magnetar
phenomena.  In this scenario FRB's are bright bursts or giant pulses
from magnetars at the centers of nearby external
galaxies, within a few hundred Mpc. 
The dominant DM contribution is due to the nuclear medium,
which is sufficient for galaxies similar to the Milky Way whose
innermost 100 pc provides $\sim$1000 pc cm$^{-3}$.

Though we do not know to what extent magnetars 
preferentially form at the GC, the fact
that one of just four radio loud magnetars is within 2.2" of Sgr A* tells us such 
objects are over represented in these environments. There are also physical 
arguments that could explain the lack of pulsars and the apparent tendency 
to form magnetars: \cite{2014ApJ...783L...7D} suggest efficient formation 
could be due to highly magnetized progenitors or a top-heavy initial mass
function. Given the large energy 
released in the episodic outbursts of SGR magnetars and the tendency for
some pulsars to emit giant pulses, we have shown that FRB's could be 
nuclear events. 
This picture also alleviates the difficulty of producing 1 ms scattering 
tails from the diffuse IGM, which has been shown to be problematic 
by \cite{2014ApJ...785L..26L} and \cite{2013ApJ...776..125M}. Though 
we do not quantify scattering from galactic nuclei, we think temporal broadening 
from such regions at $\sim0.1-100$ ms is reasonable.
Our explanation is also 
consistent with the polarization properties of FRB 140514, which had
no detectable linear polarization and $\sim$20$\%$ circular polarization. 
This could be caused by linear depolarization at low frequencies due 
to phase randomization from multiple paths through a scattering 
screen. Such polarization properties are seen in the galactic center
magnetar and giant pulses from other pulsars. 

This model is readily testable with expected upcoming surveys. With 
either a precise VLBI localization, or a large sample as expected
from UTMOST\footnote{http://www.caastro.org/news/2014-utmost} and 
CHIME \cite{2014SPIE.9145E..22B}, this model makes quantitative 
predictions.

\section{Acknowledgements}

We thank NSERC for support.

\bibliography{frbpenconnor_2april}
\bibliographystyle{emulateapj}

\label{lastpage}

\end{document}